# Formation of the Moon and Binary Asteroids

Nick Gorkavyi

**Abstract**
The proposed multi-impact model explains the formation of the Moon, Charon, and binary asteroids without invoking catastrophic cosmic events. The main elements of the new model are as follows: 1. A primordial, low-mass proto-satellite disk with prograde rotation existed around the proto-Earth. 2. Most of the lunar material was ejected from Earth's mantle by numerous impacts of large asteroids. This naturally explains the Moon's iron deficiency. 3. Collisions between Earth's ejecta and particles of the prograde proto-satellite disk stabilize the fragments on satellite orbits. We demonstrate the high efficiency of the multi-impact mechanism: Earth ejecta on prograde orbits readily merges with the prograde proto-satellite disk, whereas retrograde ejecta falls back onto Earth.

## I. Introduction

The problem of the Moon's origin has hovered over human thought since time immemorial. Descriptions of the first scientific models of lunar cosmogony can be found in the books by Ruskol and Ringwood (Ruskol, 1975; Ringwood, 1979). Ruskol developed an accretional theory of the Moon's formation from a near-Earth proto-satellite disk assembled from particles captured from heliocentric orbits (Ruskol, 1975). Such a model is applicable to the formation of a number of satellites in the Solar System, but in the case of the Moon it encounters two serious difficulties: (a) the inefficiency of the accretion mechanism in explaining the Moon's unusually large relative mass (1/81 of the planet's mass, which in 1975 was a record for the Solar System); (b) the puzzling chemical composition of the Moon, for example its low mean density (3.3 g/cm³) and its iron deficiency (13% FeO). According to the accretional theory, most researchers expected the Moon to be similar to Earth, which has a density of 5.5 g/cm³ and an average iron content of about 31% FeO.

In 1975, Hartmann and Davis proposed a unique explanation for the Moon's unusually large mass: its formation by a giant impact, in which a Mars-sized body struck the Earth along a near-tangential trajectory. Even larger impactors are now being considered; in modern simulations, the mass of the colliding body can reach about half the mass of the proto-Earth (Cameron, 2000). One consequence of this catastrophe was the formation of a massive proto-lunar disk with a large angular momentum. In this scenario, the disk—and subsequently the satellite—formed from Earth's mantle, which has the same density as the Moon. This is thought to resolve the problem of the Moon's iron depletion, since Earth's crust and mantle are relatively poor in iron (about 8% FeO) due to its concentration in the molten terrestrial core, where FeO reaches about 85%.

Over the past 30 years, the "giant impact" model has gained widespread acceptance, although at the same time a growing body of evidence has accumulated that contradicts this model (see the volume *The Origin of the Earth and the Moon*, 2000).

First, Pluto's satellite Charon—discovered in 1978 and found to have a mass equal to one-eighth that of the planet—demonstrated that there is nothing unique about a massive Moon. Could a catastrophic—and rather unlikely—mechanism of satellite formation really have occurred for two planets of the Solar System at once? Even if one answers this question in the affirmative (Canup, 2004), there remains the problem of a completely non-unique, but rather widespread (on the order of 10%), binarity among small solid-surface bodies—asteroids and trans-Neptunian objects. The existence of satellites and even entire satellite systems around such low-mass bodies is one of the most remarkable and enigmatic phenomena of the Solar System, to whose study scientists from the Crimean Astrophysical Observatory (CrAO) have made a fundamental contribution (Prokofyeva et al., 1995; Prokofyeva-Mikhailovskaya, 2008). In 1994, the first satellite of an asteroid (Ida) was photographed; in 2005, the first triple asteroid (Sylvia) was discovered; and by 2006, in addition to Charon, two more small and more distant satellites had been found around Pluto. We note that applying the giant-impact model to the origin of Charon requires an extremely unlikely—especially finely tuned in terms of impact parameter—collision between planets of comparable mass (Canup, 2004).

Second, dynamical objections have been raised against the giant-impact model, based on the fact that an impact by a Mars-sized body should lead to a large eccentricity of Earth's orbit (Boyarchuk et al., 1998). In response, proponents of the giant-impact hypothesis offer only qualitative arguments about the possible dissipative reduction of the planet's eccentricity within a planetesimal disk.

Third, geochemists have opposed the giant-impact theory with numerous arguments. From the catastrophic giant-impact model, the following geochemical conclusions can be drawn (Jones and Palme, 2000):
a. The Moon should have approximately the same chemical composition as Earth's mantle;
b. The Moon should not have a core;
c. A Moon formed from Earth's material should be younger than Earth;
d. A consequence of the giant impact should be the melting of both the Moon and Earth; thus, both bodies should have possessed an ocean of liquid magma;
e. The depletion of the Moon in volatile elements should be a result of the heating of terrestrial material during the giant impact.

In reality, however, geochemical data indicate a different picture (Jones and Palme, 2000):
a. The chemical composition of the Moon differs noticeably from that of Earth. In particular, the iron concentration in the Moon turned out to be 1.5–2 times higher than in Earth's mantle;
b. The Moon has a substantial core (1–3% by mass, or with a radius of 300–400 km, which is about 20% of the Moon's radius of 1738 km) (see also Hood and Zuber, 2000);
c. The Moon appears to be older than Earth, or more precisely, the Moon's core formed earlier than Earth's core;
d. The Moon was relatively cool and experienced only partial flooding by magma (see also Pritchard and Stevenson, 2000; Snyder et al., 2000). Geochemical data also rule out the existence of a molten-mantle ocean on Earth. For example, Earth's present-day mantle is differentiated much more weakly than it would be if an ancient global magma ocean had existed;
e. The abundance of volatiles on the Moon does not support the giant-impact model—their observed levels cannot be derived from Earth's mantle by heating.

Under the pressure of geochemical arguments, a number of scientists (Jones and Palme, 2000) have spoken in favor of the Moon's formation from a near-Earth protosatellite disk created by an accretional mechanism rather than by a giant impact.

Spudis notes that the giant-impact model has too many free parameters, which sharply reduces its scientific and predictive value (Spudis, 1996). Arguments against the giant impact are also discussed in a review by Stewart (Stewart, 2000). Nevertheless, the giant-impact model has demonstrated its success or at least adequacy in addressing a number of problems related to the Moon's formation, has gained adherents, and has acquired the inertia of a paradigm.

## 2. The Multi-Impact Model

The author (Gorkavyi, 2004) considered a new approach to the formation of the Moon, Charon, and binary asteroids: a multi-impact model which, while preserving all the achievements of the giant-impact model, eliminates its catastrophic nature, low probability, and geochemical problems. The multi-impact model is in many respects close to the classical accretional theory of lunar formation, addressing the problems of its insufficient efficiency and the mismatch between the chemical compositions of the Moon and Earth. The idea of not a single giant impact, but many "macro-impacts" in the formation of the Moon, had previously been mentioned by Ruskol (Ruskol, 1986).

The main feature of the proposed model is the rejection of the assumption of a single giant impact and the transition to the concept of numerous smaller collisions that nevertheless operate in a similar way, ejecting mantle material from the planet onto circumplanetary orbits. Naturally, the question arises as to how this material can be retained in orbit: according to Kepler's laws and the two-body problem, ejecta launched onto orbits with eccentricity $e > 1$ escapes forever, while for eccentricities $e < 1$ it must fall back onto the planet within one orbital period. Precisely to resolve this problem, the giant-impact theory requires the presence of a Mars-sized impactor, massive enough to violate the conditions of the two-body problem and thereby allow a small fraction of the ejected material to be retained on stable satellite orbits.

In the multi-impact model, the problem of orbital stability is addressed by a second fundamental element of the system—a low-mass seed protosatellite disk around the planet, the formation of which was examined in detail within the accretional model (Ruskol, 1975). Thus, the bombardment of Earth by millions of large asteroids, 10–1000 km in size, and the ejection of enormous amounts of material into space cannot by itself form the Moon, since all ejected particles either fall back onto the planet or escape onto heliocentric orbits. A protosatellite disk formed from particles on heliocentric orbits also cannot produce the observed Moon because of its low mass. We therefore consider the combined efficiency of these two factors, which are ineffective when acting separately, and show that their interaction leads to a new, successful mechanism for the Moon's formation—one that possesses the positive features of both "parent" models while avoiding their shortcomings.

Let us consider the following planar problem for the case of Earth:
- The elliptical trajectory of an ejecta particle intersects the circular orbit of a satellite particle at the ascending branch (point A) and at the descending branch (point B)—see Fig. 1.
- The semi-major axis and eccentricity of the intersecting orbits are specified; therefore, at points A and B we can readily determine the azimuthal and radial velocity components of each particle (the radial velocity component of the satellite particle is zero).
- The collision of the two particles produces a debris cloud whose center of mass moves away from the collision point with velocities determined by the law of momentum conservation (for a given ratio of the mass of the ejecta particle to the mass of the satellite particle).
- Knowing the velocity of the debris center of mass, we determine the semi-major axis and eccentricity of its trajectory and analyze:
  a. Do the fragments fall onto the planet's surface, or do they remain on a stable satellite orbit?

b. How does the orbit of the fragments change as a function of the mass ratio, the parameters of the initial ejecta orbit, and the radius of the circular orbit of the satellite particle?

c. What is the integrated effect of the interaction between a continuous flux of ejecta and the protolunar disk? Does the disk retain its orbital stability? Does the disk lose or gain mass under the influence of ejecta from the planet?

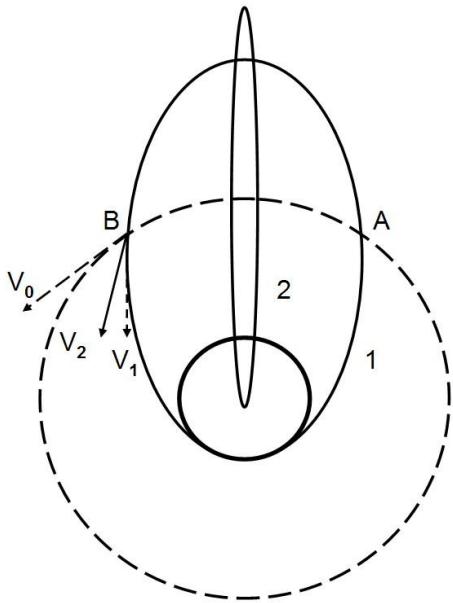

Fig. 1. Two ejecta orbits (1 and 2) intersecting the circular orbit of a particle of the protosatellite disk (dashed line). Orbit 1 is tangent to the surface, while orbit 2 corresponds to ejecta launched nearly vertically from the planet's surface. When an ejecta particle with velocity $V_1$ collides with a disk particle with velocity $V_0$, a debris cloud is formed whose center-of-mass velocity is $V_2$.

All equations describing the above processes are analytical and can be readily solved within the framework of the two-body problem (the influence of the Sun and the effect of Earth's rotation are neglected). As can be easily seen from the equations for the semi-major axis and eccentricity (Roy, 2004), these orbital parameters of the debris depend only on the square of the particle's velocity; therefore, they are exactly the same at points A and B.

What happens when an ejecta particle is launched into the vicinity of Earth surrounded by a prograde-rotating protosatellite disk? There are three main scenarios for the fate of ejecta with $e < 1$:

1. The ejecta does not interact with disk particles and therefore falls back onto Earth;
2. Ejecta on a retrograde orbit interacts with particles of the prograde disk and therefore inevitably falls back onto Earth, carrying along a disk particle if its mass is smaller than or comparable to that of the ejecta particle;
3. Ejecta on a prograde orbit interacts with particles of the prograde disk and therefore, with high probability, transitions onto a stable satellite orbit, modifying the local density of the disk.

We assume that the ejecta is symmetric—that is, for a certain number of particles launched from the planet's surface onto prograde circumplanetary orbits, there is an equal number of particles launched onto retrograde orbits. We assume that the interaction of retrograde ejecta with a particle of the prograde-rotating protosatellite disk leads to the debris falling onto the planet; thus, retrograde ejecta results in an unconditional loss of disk mass. The interaction of prograde ejecta with disk particles leads to the destruction of a disk particle on that orbit and the transfer of the particle, together with the ejecta particle (in the form of a debris cloud), onto a new orbit.

The objectives of this work are:
- to analyze individual debris orbits depending on the initial conditions of the problem;
- to consider the distribution of hundreds of millions of debris clouds over semi-major axes and analyze the change in disk mass due to bombardment by ejecta particles and the orbital transfer of debris clouds.

## 3. Calculation Results

Let us consider ejecta with a significant eccentricity $e=0.97$, a semi-major axis of 212,600 km, and a pericenter $a(1-e)$ coinciding with the Earth's surface ($R=6{,}378$ km), which corresponds to a tangential particle ejection. Suppose the ejecta intersects the circular protosatellite disk and collides with its particles. We then obtain the orbital parameters of the resulting debris cloud, assuming two values for the mass ratio of the satellite particle to the ejecta particle: 10 and 100. That is, we are considering the extreme case of very massive ejecta particles and light satellite particles.

Obviously, the debris cloud will have an eccentricity between 0 and 0.97, and presumably a pericenter larger than the initial one due to its sensitivity to the eccentricity value. The pericenter of the debris cloud is a critical parameter—if the pericenter increases by only a few percent above the initial value, the debris will no longer fall onto Earth but will remain on a stable satellite orbit. This allows the debris, gradually through subsequent collisions with disk particles and with each other, to join the circular disk.

Figure 2 shows the pericenter of the debris cloud as a function of the radius of the circular orbit of the satellite particle involved in the collision. It can be seen that the overwhelming majority of ejecta particles, even with a mass 100 times greater than that of the satellite particle, raise their pericenters above Earth's atmosphere and remain on stable orbits. Figures 3 and 4 show the eccentricity and semi-major axis of the debris cloud as functions of the satellite particle's orbital radius.

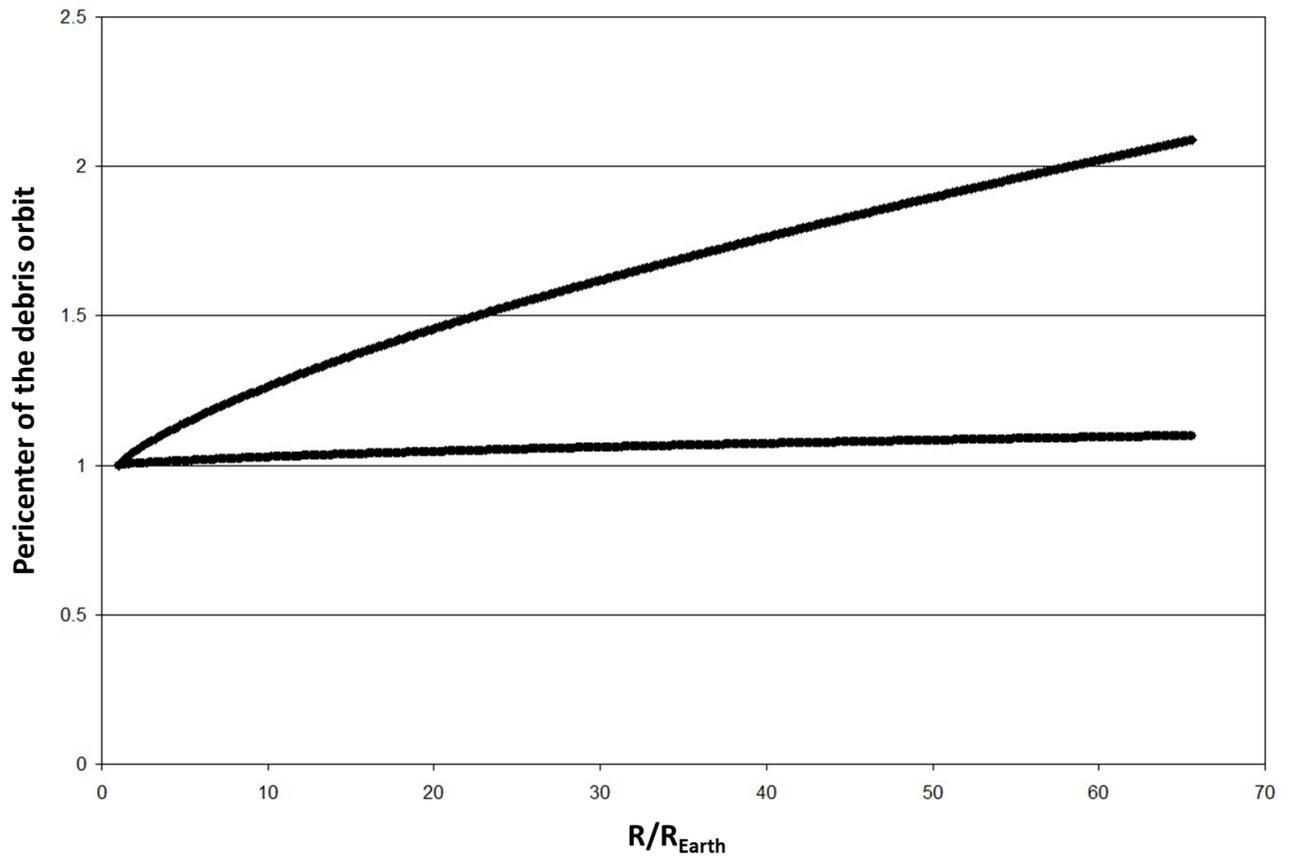

**Fig. 2.** Pericenter of the debris cloud orbit after the collision of ejecta with a disk particle as a function of the radius of the disk particle's orbit. The ejecta has a semi-major axis of 212,600 km and an eccentricity of 0.97. The upper curve corresponds to ejecta particles with a mass 10 times greater than the mass of the disk particle, while the lower curve corresponds to a mass 100 times greater.

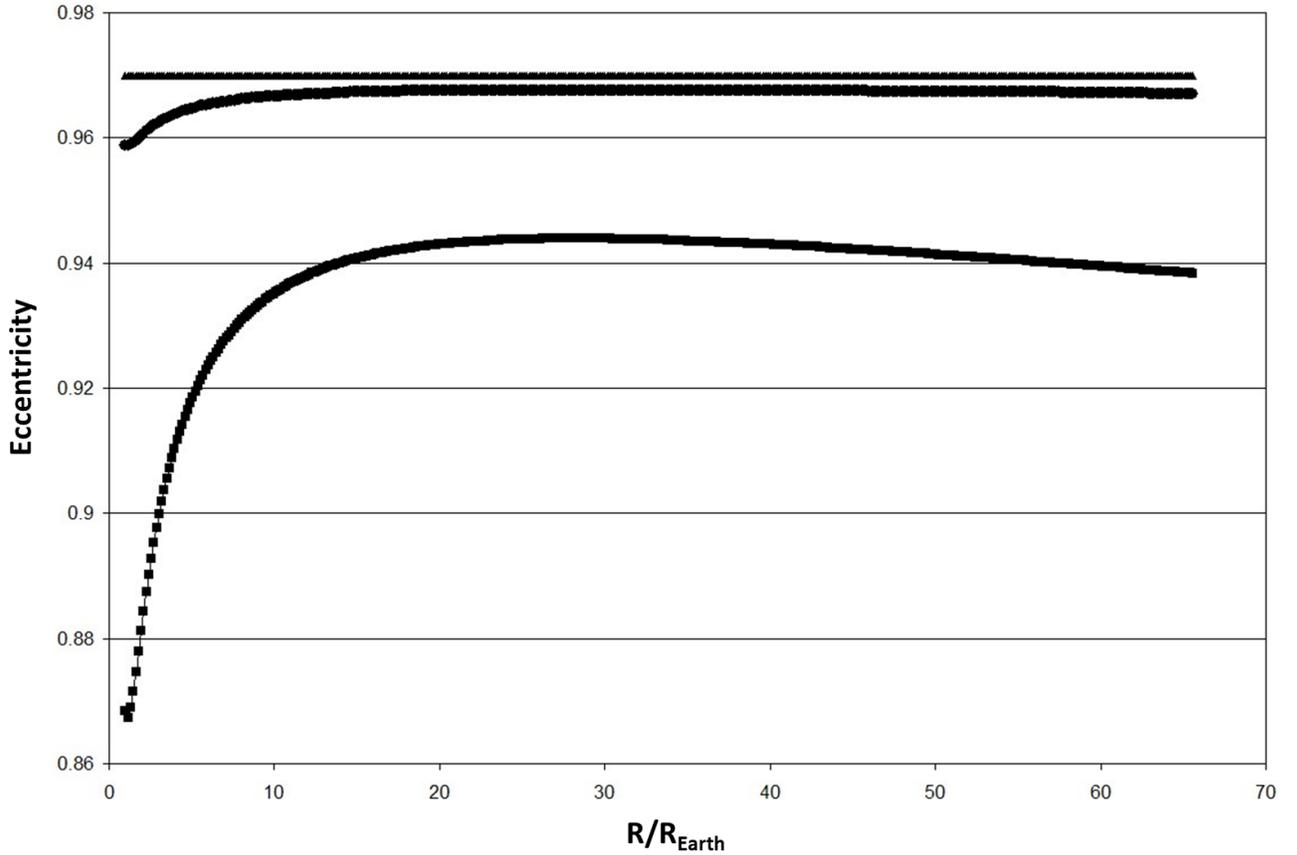

**Fig. 3.** Eccentricity of the debris cloud orbit after the collision of ejecta with a disk particle as a function of the radius of the disk particle's orbit involved in the collision. The ejecta has a semi-major axis of 212,600 km and an eccentricity of 0.97 (the uppermost straight line). The upper curve corresponds to ejecta particles with a mass 100 times greater than the disk particle, while the lower curve corresponds to a mass 10 times greater.

The semi-major axis of the debris cloud changes depending on the orbital distance to the collision point and the orbital parameters of the ejecta (Fig. 4): the ejecta repels nearby disk particles from the planet, increasing their semi-major axis, and attracts more distant particles, decreasing their semi-major axis, thereby creating a peak in the preferential concentration of debris orbits. A similar effect was analytically and numerically studied for dust ejecta in the Uranian rings (Fridman and Gorkavyi, 1999). This repulsion is enhanced if we increase the eccentricity and semi-major axis of the ejecta orbit (while maintaining tangency to Earth's surface) or the mass of the ejecta particles. This effect resolves the question of the potential collapse of the protosatellite disk onto the planet due to bombardment by ejecta: in fact, ejecta with sufficient eccentricity imparts additional angular momentum to the disk. This effect persists across wide variations in the initial parameters and is observed even when the assumption of tangency of the initial ejecta trajectory is abandoned.

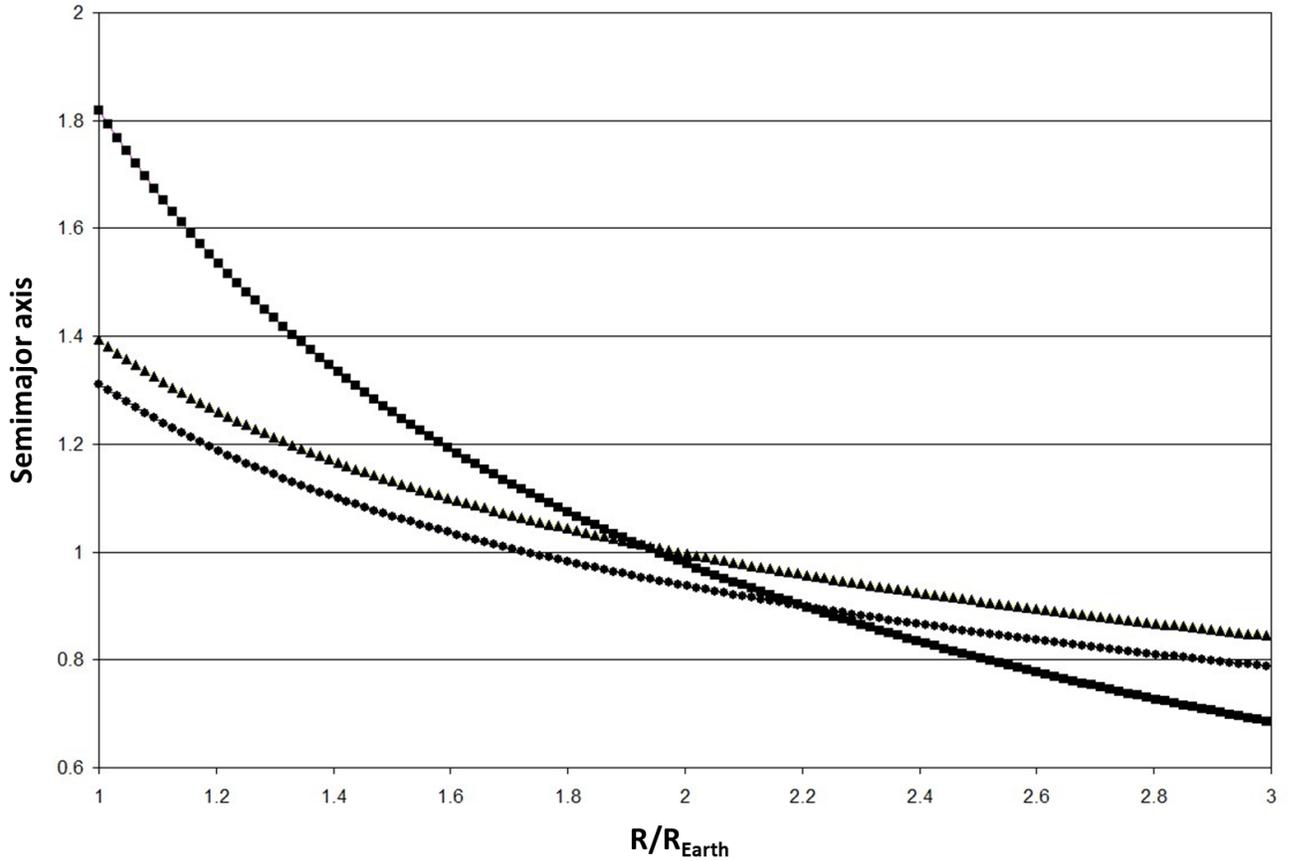

**Fig. 4.** Change in the semi-major axis of the debris cloud orbit relative to the semi-major axis of the disk particle's orbit as a function of orbital distance from the planet. No change corresponds to 1. The curve with circles corresponds to ejecta with a semi-major axis of 12,756 km, an eccentricity of 0.5, and a mass ratio to the disk particle of 1. The curve with triangles corresponds to ejecta with a semi-major axis of 15,945 km, an eccentricity of 0.6, and a mass ratio of 1. The curve with squares corresponds to ejecta with a semi-major axis of 12,756 km, an eccentricity of 0.5, and a mass ratio to the disk particle of 10 (ejecta particle is heavier).

Within the assumptions described in the previous section, we consider the evolution of a protosatellite disk of uniform surface density, with a radius exceeding Earth's radius by 100,000 km. Figure 5 shows the curves of growth and degradation of this disk due to bombardment by ejecta (both prograde and retrograde). We considered 1,000 different values of ejecta particle mass, from 0.1 to 10 relative to the mass of disk particles, taken as 1. The number of particles is inversely proportional to their mass. We considered 1,000 disk orbits (with a step of 100 km) at different radii and 400 ejecta orbits with eccentricities uniformly distributed from nearly circular to 1. All ejecta orbits are tangent to Earth's surface, so the semi-major axes correspond to this condition. The probability of interaction between ejecta and each disk orbit was assumed to be 1%. In total, 400 million trajectories of debris clouds were considered. We assume that each debris cloud contributes to the disk orbit with a radius equal to the semi-major axis of the cloud's orbit, i.e., its average distance from the planet.

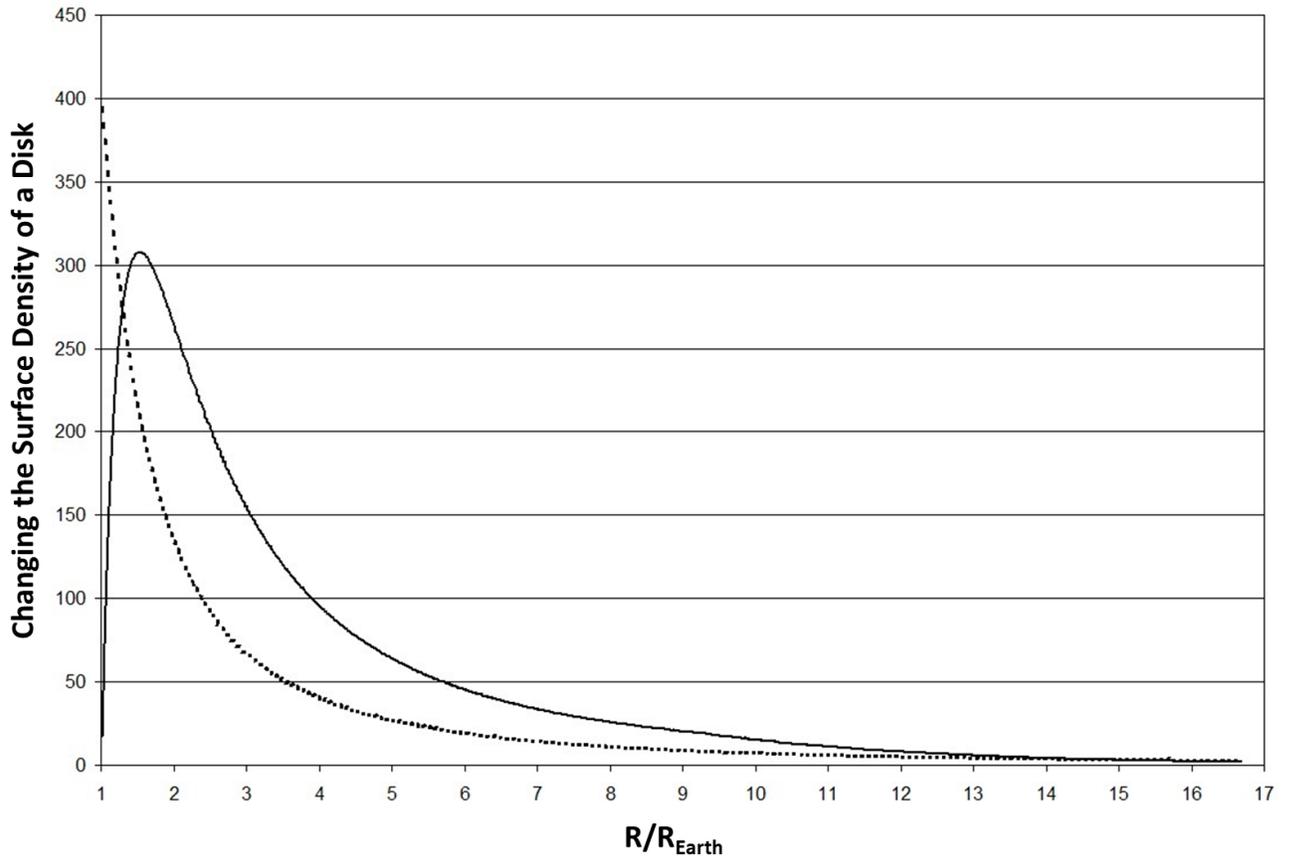

**Fig. 5.** Change in the surface density of the protosatellite disk due to interaction with ejecta launched from Earth. The initial disk, extending 100,000 km from Earth's surface, is assumed to have a uniform surface density. The dashed curve represents the rate of decrease in the disk's surface density due to interaction with ejecta (both prograde and retrograde), while the solid curve represents the rate of increase in the disk's surface density due to the accretion of prograde ejecta. Over nearly the entire disk (except for the innermost and outermost regions), the increase in disk density dominates over the decrease.

It can be seen that the parts of the protosatellite disk closest to the planet are depleted (due to outward displacement of particles and falling onto the planet), whereas in the rest of the disk there is a steady growth that dominates over depletion (except in the very outer edge of the disk). Disk growth occurs due to prograde ejecta, which compensates for the loss of disk material caused by retrograde ejecta, thereby realizing an effective selection mechanism for material ejected from the planet's surface onto prograde orbits aligned with the rotation of the protolunar disk. The results remain qualitatively unchanged under significant variations in the size distribution of ejecta particles.

The accretion rate of satellites from the protosatellite disk can be very high, and a scenario in which the disk rapidly loses its density and its ability to interact with ejecta is possible. In such a case, only a small inner part of the disk may remain around the planet, corresponding to planetary rings, whose accretion into a satellite is limited by the high collision velocities of particles within the disk (Gorkavyi and Fridman, 1994). Fig. 6 considers the case where ejecta interacts only with a small disk extending 10,000 km above Earth's surface. We see that the principal features of disk evolution—depletion of density near the planet and growth at all other distances—are preserved.

Unlike the calculation condition shown in Fig. 5, in Fig. 6 the step between 1,000 orbits of the initial disk was chosen as 10 km, and the distribution of debris clouds was performed over a space twice as large—20,000 km above Earth's surface (2,000 orbits). In total, 400 million trajectories of debris clouds were considered.

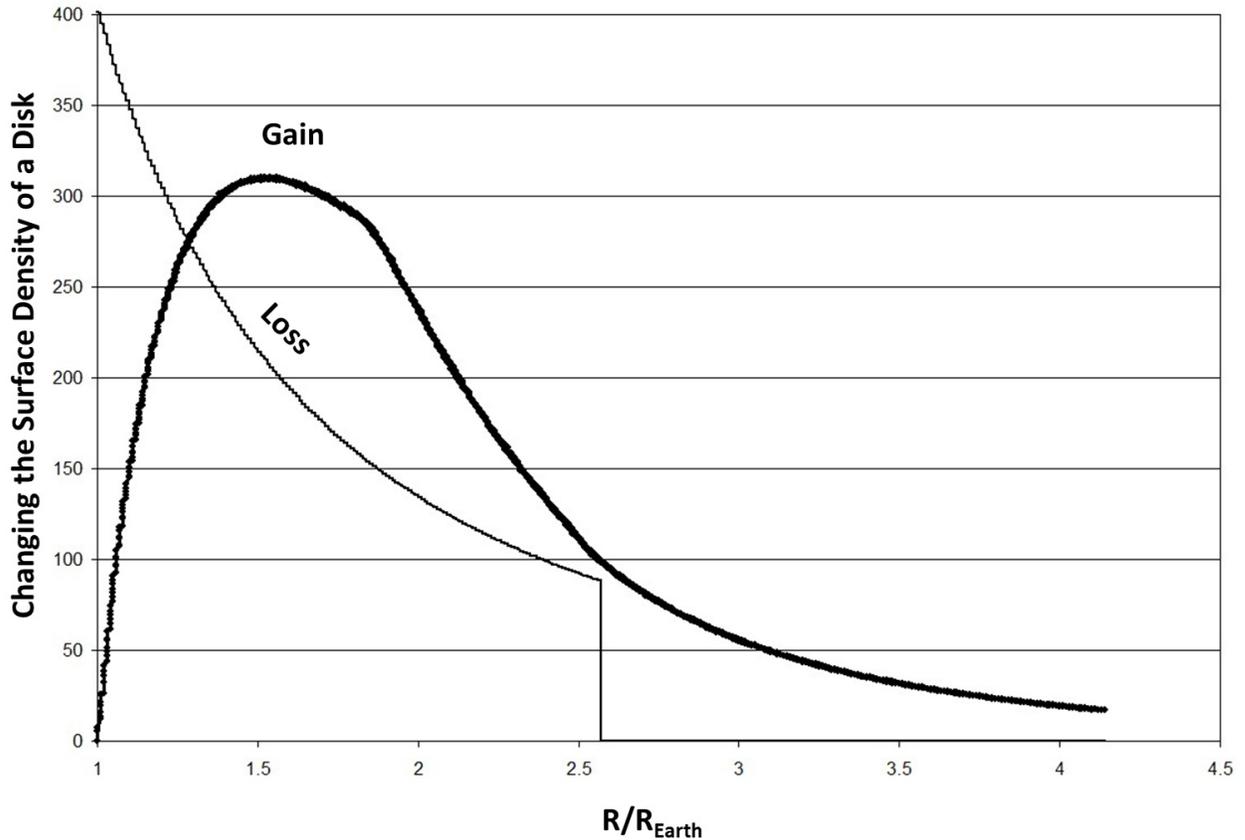

**Fig. 6.** Change in the surface density of the protosatellite disk due to interaction with ejecta launched from Earth. The initial disk, extending 10,000 km above Earth's surface, is assumed to be uniform. The thin curve represents the rate of decrease in the disk's surface density due to interaction with ejecta (both prograde and retrograde), while the thick curve represents the rate of increase in the disk's surface density due to accretion of prograde ejecta. The best conditions for satellite formation are created in the zone immediately beyond the planetary rings; in this calculation, this corresponds to a distance of 2.5–3 Earth radii.

These calculations convincingly demonstrate the dynamic efficiency of the proposed mechanism for the Moon's formation.

**4. General Scenario for the Origin of Satellite Systems in the Solar System**

From our point of view, there are two main types of satellite systems in the Solar System and two corresponding formation mechanisms. The satellite systems of different planets can be divided into two groups according to the ratio of the average orbital velocity to the escape velocity from the planet's surface:

**Giant planets**, for which the orbital velocity is 3–5 times smaller than the escape velocity:
- Jupiter – 0.22
- Saturn – 0.27
- Uranus – 0.32
- Neptune – 0.23

**Terrestrial planets**, for which the orbital velocity is 3–11 times larger than the escape velocity:
- Mercury – 11.3
- Venus – 3.4
- Earth – 2.7
- Mars – 4.8
- Pluto – 3.7

This group also includes all trans-Neptunian objects, asteroids, and comets.

From a dynamical perspective, the division of planets into two groups corresponds to different ratios between typical impact velocities of bodies striking the planet and the velocity required to eject potential ejecta from that planet. The velocities of gas and solid bodies relative to the planet are proportional to the planet's orbital velocity and, although smaller, are comparable to it.

**Satellite Systems of Giant Planets.** All material falling from heliocentric orbits onto a giant planet becomes incorporated into the planet upon any contact with its surface. Material arriving from heliocentric orbits—both gas and solid bodies—can not only fall onto the planet but also remain on satellite orbits if it interacts favorably, not with the planet, but with itself. This process of accretional formation is described in detail by Ruskol (Ruskol, 1975). The typical mass of a giant planet's satellite system is about 0.0001–0.0002 of the planet's mass and is contained in usually regular, massive satellites, such as the Galilean moons.

Small and irregular outer satellites of giant planets appear very different from the inner massive satellites, but their formation is also linked to quasi-accretional capture of particles and asteroids from heliocentric orbits. Gorkavyi and Taidakova, in a series of papers from 1993–1995 (see Gorkavyi and Taidakova, 1995), showed that the interaction of solid bodies arriving from heliocentric orbits with a protosatellite disk can explain the formation of groups of irregular satellites—both prograde, such as the Himalia and Nereid groups, and retrograde, such as the Pasiphae and Phoebe groups, and even the massive retrograde Triton.

In the 1995 paper, a prediction was made regarding the existence of an as-yet-undiscovered, outermost group of retrograde satellites of Saturn. Calculations of 256,000 debris clouds in the three-body problem (debris–Saturn–Sun) showed that the radius of the retrograde satellite zone begins at 19 million km, with a peak accumulation around 25–26 million km. In 1997–1998, the author, together with Japanese astronomers, attempted to include the search for Saturn's outer satellites in the program for the new Subaru telescope. The attempt was unsuccessful, but in 2000, the predicted group of outer retrograde Saturnian satellites was discovered by Gladman et al. using significantly smaller telescopes. By 2006, 16 retrograde satellites were discovered in the zone from 18.4 to 23 million km; the Subaru telescope has been used to observe these satellites.

Model calculations (Gorkavyi and Taidakova, 1995) predicted that the retrograde Phoebe (12.9 million km) should, conversely, be surrounded by prograde satellites. By 2006, in the zone around Phoebe—from 11.1 to 18 million km—8 prograde satellites and only one retrograde satellite had been discovered. Gorkavyi and Taidakova (2002) discussed in January 2002 the agreement between theoretical and observational results for Saturn's irregular satellites and, based on the 1995 model, made the additional prediction that there is a group of numerous undiscovered satellites beyond Triton's orbit around Neptune (>0.5 million km). It was noted that in this zone, prograde satellites should be mixed with more numerous retrograde satellites. The discovery of this group of Neptune satellites by astronomers under Holman–Cavelears was announced a few months later in 2002. By 2006, in the zone from 15.7 to 48.4 million km, 2 prograde and 3 retrograde satellites of Neptune had been discovered.

Thus, the general course of satellite system formation around giant planets is well understood and is accurately described by the accretional model in its various forms.

**Satellite Systems of Terrestrial Planets and Asteroids.** For small, solid-surfaced planets, a different mechanism of satellite formation operates, in which, in addition to classical accretion, the growth of the disk is largely determined by the flux of material ejected from the planet's surface. Consider, for example, the extreme case of a small asteroid. Its orbital velocity and the relative velocities of meteoroids in its vicinity are several kilometers per second, far exceeding the speed of bullets or shells. In contrast, the orbital velocities of satellites around the asteroid are only a few meters per second. To capture incoming material onto its satellite orbit, the asteroid must reduce the velocity of the incoming particle by thousands of times.

This can occur in only one way—by presenting its surface for the particle to collide with. A micrometeoroid striking the loose regolith on the asteroid's surface dissipates its velocity and transfers all its energy to the surrounding material, which is much more massive than the incoming meteoroid. Part of the regolith will be lost into space, but some of the ejecta will acquire moderate velocities and, after interacting with the cloud or the protosatellite particle disk around the asteroid, will remain on a stable satellite orbit, thereby increasing the mass of the protosatellite cloud.

The asteroid itself may even be partially destroyed during such bombardment, but by sacrificing some of its own material, it grows a satellite disk large enough to form a quasi-spherical satellite on a circular orbit near the asteroid's equator—in accordance with the principles of accretion theory.

Subsequent micrometeoroid bombardment of the asteroid then leads to growth not of the disk, but of the already formed satellite, and, due to the symmetry of prograde and retrograde ejecta, to braking of the orbital motion of the asteroid's components. Under certain conditions, the satellite may slow down and merge with the primary body, forming a dumbbell-shaped object. Material exchange between the components of a binary asteroid can also lead to comparable masses. The formation of an additional satellite on an outer orbit around such a binary is quite plausible. The third body can grow and, undergoing a similar evolution, slow down and join the already merged pair of bodies. This likely explains the formation of the triple asteroid 9969 Braille (see Prokofyeva-Mikhailovskaya, 2008). Strong external impacts can also increase the satellite's eccentricity, as observed for some trans-Neptunian objects. Rapid rotation of the central body should promote the growth of a prograde satellite disk and the successful formation of a satellite. Observations confirm a correlation between the rotation rate of the central body and the presence of satellites, as well as the regular nature of asteroid satellite systems, in particular the prograde rotation of their satellites (Gaftonyuk and Gorkavyi, 2013).

We propose that Mars' satellite system—Phobos and Deimos—formed similarly to asteroid satellites. Phobos is located at a distance of 2.76 Mars radii—practically very close to the zone of planetary rings and the region of most efficient ejecta accumulation (see Fig. 6). Note that with Mars' density of 3.94 g/cm³, Phobos has a density of 1.9 g/cm³, and the more distant Deimos 1.75 g/cm³. Typically, satellites of giant planets are denser than their gaseous hosts, whereas satellites of terrestrial planets, if large enough to undergo mantle differentiation, are less dense precisely due to the inflow of planetary mantle material.

The formation of the Moon and Charon differs from that of Phobos–Deimos and binary asteroids only in that these larger satellites moved significantly farther from their planets due to tidal effects. In this process, the Moon, apparently faster in terms of radial drift and longer-lived, was able to "consume" the smaller outer satellites of Earth, whereas the slower and younger Charon did not.

It is possible that the widely discussed enigmatic intense bombardment of the Moon 3.9 billion years ago, traces of which are absent on Earth (see Ryder et al., 2000), resulted from collisions of outer, smaller satellites with the large Moon drifting outward. Several authors (Ruskol, 1975; Spudis, 1996) discuss the problem of asymmetry in craters and maria on the near and far sides of the Moon. Perhaps the greater number of maria on the Moon's near side (and especially on the leading and visible quarter of the lunar disk) indicates bombardment of our satellite by ejecta originating from Earth. Thus, the "mare" asymmetry points to multiple and temporally distributed strong asteroid impacts that transferred material from Earth to the Moon.

## 5. Conclusion

The goal of this work was not to construct a detailed quantitative model of the formation of the Moon and binary asteroids, but to demonstrate the effectiveness of the mechanism for accumulating massive satellite disks from the mantle material of terrestrial planets and small planets. This goal has been achieved: we have shown that:
- Ejecta launched from a planet's surface easily transitions to stable satellite orbits, even when interacting with small protosatellite disk particles that are 10–100 times less massive than the ejecta particles;
- A prograde-rotating protosatellite disk efficiently captures and incorporates prograde ejecta, while destabilizing and returning retrograde ejecta to the planet;
- The initial protosatellite disk or planetary rings are not only not destroyed by the action of oppositely directed ejecta, but, remaining stable, actively grow due to prograde ejecta, which contributes mass and additional angular momentum to the disk.

The model considered here is a natural extension of the giant-impact idea, integrating it with the classical accretion model. The giant-impact concept made an exceptional contribution to understanding the Moon's formation by introducing the crucial thesis of the importance of mantle material flux caused by collisions of large planetesimals with the planet. The accretion model addressed the formation and evolution of the second key element in the cosmogony of the Moon—the protosatellite disk around Earth.

It is now time to combine these two models—accretion and giant-impact—into a unified, effective multi-impact theory, applicable to modeling the formation of a wide range of satellite systems, from large planets like Earth to small asteroids only a few hundred meters across. The renewed surge of interest in lunar studies at the start of the new millennium allows hope that, given the growing information about the Moon and the practical needs of astronauts who will explore Earth's satellite, scientists from different schools of thought will be able to find common ground on the longstanding problem of the formation of our night-time companion.